\documentclass[twocolumn,showpacs,preprintnumbers,amssymb,amsmath,prl]{revtex4}
\usepackage{epsfig}
\usepackage{dcolumn}

\begin{document}

\bibliographystyle{apsrev}
\preprint{code number:}
\title{Probing interface magnetism in the FeMn/NiFe exchange bias
system using  magnetic second harmonic generation }

\author{Luiz C. Sampaio}
\altaffiliation[On leave from]{ Centro Brasileiro de Pesquisas
Fisicas Rua Dr. Xavier Sigaud, 150 Urca, Rio de Janeiro, RJ
22.290-180}
\author{Alexandra Mougin}
\email{mougin@lps.u-psud.fr}
\author{Jacques Ferr\'e}
\affiliation{Laboratoire de Physique des Solides, UMR CNRS 8502,
B\^at.~510, Universit\'e Paris Sud, 91405 Orsay, France}

\author{Patrick Georges}
\affiliation{Laboratoire Charles Fabry de l'Institut d'Optique
Th\'eorique et Appliqu\'ee (IOTA), B\^at.~503, UMR CNRS 8501,
Universit\'e Paris Sud, 91405 Orsay, France}
\author{Alain Brun} \affiliation{Laboratoire Charles Fabry de l'Institut
d'Optique
Th\'eorique et Appliqu\'ee (IOTA), B\^at.~503, UMR CNRS 8501,
Universit\'e Paris Sud, 91405 Orsay, France}

\author{Harry Bernas}
\affiliation{Centre de Spectrom\'etrie Nucl\'eaire et de
Spectrom\'etrie de Masse, UMR CNRS 8609, B\^at.~108, Universit\'e
Paris Sud, 91405 Orsay, France}

\author{Stefan Poppe}
\affiliation{Fachbereich Physik and Forschungs- und
Entwicklungsschwerpunkt Materialwissenschaften,
Erwin-Schr\"{o}dinger-Stra{\ss}e 56, 67663 Kaiserslautern,
Germany}

\author{Tim Mewes}
\affiliation{Fachbereich Physik and Forschungs- und
Entwicklungsschwerpunkt Materialwissenschaften,
Erwin-Schr\"{o}dinger-Stra{\ss}e 56, 67663 Kaiserslautern,
Germany}

\author{J\"urgen Fassbender}
\affiliation{Fachbereich Physik and Forschungs- und
Entwicklungsschwerpunkt Materialwissenschaften,
Erwin-Schr\"{o}dinger-Stra{\ss}e 56, 67663 Kaiserslautern,
Germany}

\author{Burkard Hillebrands}
\affiliation{Fachbereich Physik and Forschungs- und
Entwicklungsschwerpunkt Materialwissenschaften,
Erwin-Schr\"{o}dinger-Stra{\ss}e 56, 67663 Kaiserslautern,
Germany}

\date{\today}
%
%
\begin{abstract}
 Second harmonic generation magneto-optic Kerr effect (SHMOKE) experiments,
sensitive to buried interfaces, were performed on a
polycrystalline NiFe/FeMn bilayer in which areas with different
exchange bias fields were prepared using 5 KeV He ion irradiation.
Both reversible and irreversible uncompensated spins are found in
the antiferromagnetic layer close to the interface with the
ferromagnetic layer.  The SHMOKE hysteresis loop shows the same
exchange bias field as obtained from standard magnetometry. We
demonstrate that the exchange bias effect is controlled by pinned
uncompensated spins in the antiferromagnetic layer.
\end{abstract}

\maketitle

The magnetic exchange interaction between an antiferromagnetic
(AF) and an adjacent ferromagnetic (F) layer may lead to the
exchange bias effect discovered  in
1956~\cite{Meiklejohn1,Meiklejohn2}. Among other various
intriguing features, this effect leads to a shift of the F
hysteresis loop along the field axis by the so-called exchange
bias field $H_{eb}$. For recent reviews see
Refs.~\cite{Nogues,Berkowitz,Stamps}.  Proposed models to account
for the exchange bias  involve (i) domain walls or partial domain
walls in the AF layer which are either
parallel~\cite{Stamps,Mauri} or perpendicular~\cite{Malozemoff} to
the interface, and/or (ii) uncompensated AF layer magnetic moments
at the interface~\cite{Malozemoff,Schulthess,Nowak} and/or in the
bulk~\cite{Nowak,Nowak2}. In most exchange bias models, the
interfacial uncompensated spins are linked to roughness,
structural defects, or disoriented grains. Although uncompensated
spins have been already evidenced~\cite{Antel}, their behavior
during the F layer magnetic reversal has not been reported so far
and, experimentally, the relationship between uncompensated spins
and exchange bias is still unclear. In the special case where
artificial random defects can be introduced in the AF layer (such
as in a diluted antiferromagnet), the so-called ``domain state
model''~\cite{Nowak,Nowak2}  showed that the exchange bias effect
stems from the volume AF spin arrangement triggered by non
magnetic defects. In this model, AF interfacial reversible
 and irreversible uncompensated spins (creating $M_{rev}^{F/AF}$ and $M_{irr}^{AF}$ respectively)
are distinguished. Some of the interfacial AF uncompensated spins
reverse under the action of an external magnetic field and the
additional effective interface exchange field originating from the
magnetized F layer, whereas the rest of the AF uncompensated spins
remain frozen in the same range of applied fields. The reversible
uncompensated spins hysteresis loop is found to be shifted along
the field axis by $H_{eb}$ and along the magnetization axis by an
amount directly proportional to $M_{irr}^{AF}$, which scales with
$H_{eb}$~\cite{Nowak2}. Using superconducting quantum interference
device magnetometry, this vertical shift of the hysteresis loop of
F/AF bilayers has already been measured and related to the
exchange bias field sign~\cite{Nogues2}, although its origin was
not determined.

 Here we study the second-harmonic magneto-optic
Kerr effect (SHMOKE) in an exchange-bias system. A second-harmonic
signal in centrosymmetric materials is selectively generated at
their interfaces due to symmetry breaking, so that the effect only
senses about 2 monolayers in the vicinity of flat
surfaces~/~interfaces~\cite{sh,Wierenga}. In contrast, the
standard linear MOKE signal originates mainly from the bulk of the
F layers. Generally, the second harmonic optical polarization
$\vec{P}(2\omega)$, generated at a single interface, consists of
both magnetic and non-magnetic contributions due to the magnetic
optical susceptibility  $\chi _{m}(2\omega)$ (linear with respect
to the magnetization $\vec{M}$) and the non-magnetic one
$\chi_{nm}(2\omega)$ (independent or even with respect to
$\vec{M}$),

\begin{subequations}
\begin{eqnarray}
P_{i}(2\omega) &=&\sum_{j,k} \ \chi _{ijk}(2\omega) \
E_{j}(\omega)E_{k}(\omega)\\ \chi(2\omega,\vec{M})& = &\chi
_{nm}(2\omega,\vec{M}) + \chi _{m}(2\omega,\vec{M}) \label{a}\\
\chi(2\omega,-\vec{M}) &= &\chi _{nm}(2\omega,\vec{M}) - \chi
_{m}(2\omega,\vec{M}) \label{b}
\end{eqnarray}
\end{subequations}
where $E_{j}(\omega)$ are the electric field components of the
incident light and $\chi _{ijk}(2\omega)$ the second harmonic
susceptibility tensor elements. Each surface or interface
contributes to the SHMOKE signal so that the measured intensity
from a $n$ multilayer is given by the sum of all interfering
signals: $I(2\omega) \propto \sum_{(a,b)=(1,...,n)} P_{a}(2\omega)
P_{b}^{*}(2\omega)$.\\

\begin{figure}[!h]
\includegraphics[width=7cm]{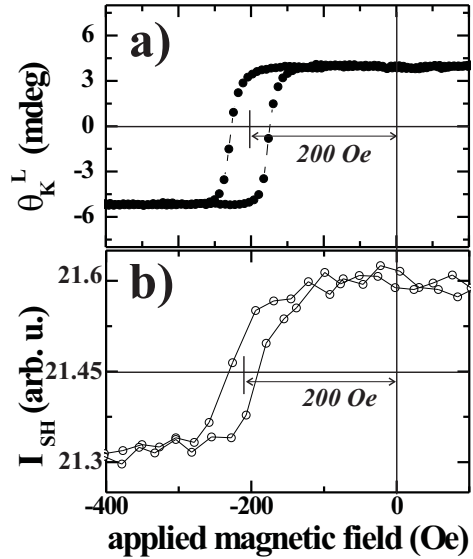}
\caption{Bulk and interfacial magnetization reversal loops
investigated by linear MOKE  (a), and second harmonic MOKE  (b),
respectively. $\theta_K^L$ is the linear Kerr rotation and
$I_{SH}$ is the SHMOKE intensity in the transverse configuration.}
\label{fig1}
\end{figure}

\begin{figure}[!h]
\includegraphics[width=8cm]{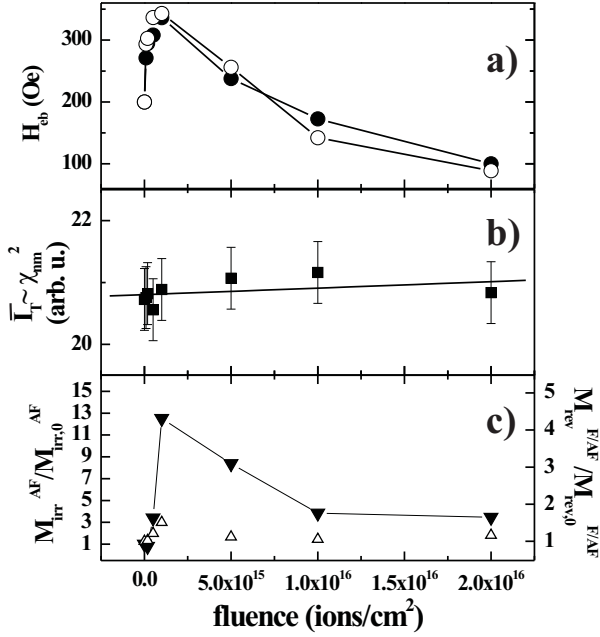}
\caption{Ion fluence dependence of: a) exchange bias field (full
symbols: MOKE, open symbols: SHMOKE), b) non-magnetic contribution
to the SHMOKE signal ($\overline{I_T}$), and c)
 irreversible
($M_{irr}^{AF}/M_{irr,0}^{AF}$, full triangles) and
  reversible
magnetization ($M_{rev}^{F/AF}/M_{rev,0}^{F/AF}$, empty
triangles), normalized to their initial value and deduced from
asymmetry measurement in the longitudinal configuration. Solid
lines are guides to the eye. } \label{fig2}
\end{figure}

In the present study, polycrystalline bilayers of ferromagnetic
Ni$_{81}$Fe$_{19}$ and antiferromagnetic Fe$_{50}$Mn$_{50}$ were
used in order to tailor the exchange bias field by light ion
irradiation~\cite{IEEE,Mougin}. Bilayers of 10~nm FeMn and 5~nm
NiFe were evaporated on a 15~nm thick Cu buffer layer deposited on
a thermally oxidized Si substrate. A thin Cr cover layer protected
the samples from oxidation. In order to initialize the
unidirectional anisotropy, the samples were heated and then field
cooled. This led to a homogeneous exchange bias field $H_{eb}$ of
200~Oe across the sample, as determined from linear MOKE (see
Fig.~\ref{fig1}a). After preparation, the exchange bias was
modified in different areas of the sample using 5~keV He ion
irradiation in a fluence range between 9$\cdot$10$^{13}$ and
2$\cdot$10$^{16}$ ions/cm$^{2}$, leading to $H_{eb}$ values
between 100~Oe  and 350~Oe (see Fig.~\ref{fig2}a, full symbols),
consistent with previous work~\cite{Mougin}. A discussion of the
exchange bias field evolution with ion fluence is given in
Ref.~\cite{Mougin}. Based upon the latter and ion stopping
calculations~\cite{ TRIM} we note that the interface roughness is
not affected when ion fluences are below a few times 10$^{16}$
ions/cm$^2$.

SHMOKE experiments were performed with a mode-locked Ti:Sapphire
laser operating at a central wavelength of 800~nm, emitting light
pulses of width 100~fs at a repetition rate of 86~MHz. All
experiments were performed in reflection; the average power at the
sample surface was 50~mW within a focus of 30-40~$\mu$m. Using a
$P$-polarized laser beam ($P_{in}$) at an incident angle of
45$^{\circ}$ the SH signal can be measured in the transverse
configuration ($T$), and in the longitudinal one ($L$) when
combined with a polarization analysis.  The complex effective
magnetic and non-magnetic susceptibilities $\chi_{m}$ and  $\chi
_{nm}$ which enter the calculation of the SHMOKE intensity are
functions of $\chi_{ijk}(2\omega)$ tensor elements depending on
the measurement geometry, hence they differ in the transverse or
longitudinal configurations~\cite{sh}. In order to extract
information on the magnetization at the F/AF interface, the SH
contribution of each individual interface and surface (Cu/FeMn,
FeMn/NiFe, NiFe/Cr and Cr/air) must be analyzed. Due to the same
crystallographic structure and the close chemical nature of the
FeMn and NiFe layers, independent of the measurement
configuration, $\chi_{nm}$ originating from the FeMn/NiFe
interface has a smaller value than those of the upper NiFe/Cr and
Cr/air interfaces. Therefore (see definition of
$R_{L}=|\frac{\chi_{m}}{\chi_{nm}}|_L$ in Eq.~(\ref{AL})),  most
of the magnetic SH signal originates from the FeMn/NiFe interface
whose $R_{L}$ is large. No signal is expected from the bulk
centrosymmetric F or AF layers~\cite{sh}.

In  transverse geometry, with $P$-polarized incident light,  the
SH outgoing beam is still $P$-polarized. Thus, the second harmonic
intensity was measured as a function of the applied field, which
was oriented perpendicular to the plane of incidence and parallel
to the exchange bias direction. The resulting hysteresis loop is
shifted along the field axis by the same bias as that of the bulk
F layer
 (Figure~\ref{fig1}b). We conclude that SHMOKE selectively probes
reversible ($M_{rev}^{F/AF}$) uncompensated spins at the
interface, which are coupled to the F layer~\cite{Nowak,Nowak2}.
Moreover, due to some frustration at the F/AF interface, the
SHMOKE hysteresis loop is broader than its bulk F layer
counterpart. As shown in Fig.~\ref{fig2}a, at all fluences the
exchange bias field values measured via SH or linear MOKE agree
within the error bars.

Another magnetization term ($M_{irr}^{AF}$), required to induce
any $H_{eb}$ and which remains unaffected by the F layer reversal
(between $\pm H$, with $|H|=H_{sat}^F$), has been probed as
follows. In order to separate out the magnetic and  non magnetic
contributions (see Eq.~(1)), the ion fluence dependence of the
non-magnetic contribution was first determined.
Commonly~\cite{Hohlfeld} since the magnetic term is small as
compared to the average part of the SH
intensity~\cite{SHintensity}, the non-magnetic contribution scales
with the average transverse SHMOKE intensity
$\overline{I_T}$~\cite{Gudde}:

\begin{subequations}
\begin{eqnarray}\label{average}
\overline{I_T} &=& \frac{I_T(2\omega,+H)+I_T(2\omega,-H)} {2} \\
&\propto& |\alpha \chi_{nm}|_T^2 + |\beta \chi_{m}|_T^2  \approx
|\alpha \chi_{nm}|_T^2
\end{eqnarray}
\end{subequations}

\noindent $\alpha$ and $\beta$ are the effective Fresnel
factors~\cite{Gudde}. $\overline{I_T}$ (and thus the effective
non-magnetic optical susceptibility $|\chi_{nm}^2|_T$) is shown in
Fig.~\ref{fig2}b as a function of ion fluence. Practically no
evolution is observed,  confirming that no significant interface
broadening occurs upon irradiation in this fluence range.

\begin{figure}[!h]
\includegraphics[width=7cm]{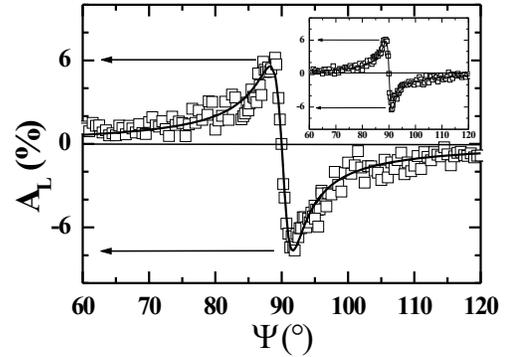}
\caption{SHMOKE asymmetry $A_L$ in longitudinal geometry as a
function of the analyzer angle $\psi$ at a fluence of 10$^{15}$ He
ions/cm$^2$ (10$^{16}$ He ions/cm$^2$ in the inset). The solid
line is the fit described in the text. \label{fig3}}
\end{figure}

Polarization measurements were performed in the longitudinal
configuration, and the magnetic contributions to the SH signal
were extracted from the asymmetry
 $A_{L}(H,\psi)$, defined as the normalized intensity
difference when the F magnetization is reversed \cite{sh}. At the
F/AF interface, the structural ($\chi _{nm}$) and two distinct
magnetic ($\chi_{m}(M_{rev}^{F/AF})$ and $\chi_m(M_{irr}^{AF})$)
contributions can give rise to a second harmonic signal. Each
magnetic contribution induces a rotation of the polarization of
the outgoing beam (from $P$ to $S$)~\cite{sh} and is phase shifted
relative to the structural term: $\varphi_{rev}$ (respectively
$\varphi_{irr}$) is the phase angle between
$\chi_m(M_{rev}^{F/AF})$ (respectively $\chi_m(M_{irr}^{AF})$) and
$\chi_{nm}$. Usually, when the applied field is reversed, all
magnetic contributions (linear with M) to the second harmonic
polarization change sign. Here, between $\pm H$, we assume that
$M_{rev}^{F/AF}$ reverses i.e. $\chi_{m}(M_{rev}^{F/AF})$ changes
its sign whereas $M_{irr}^{AF}$ is pinned i.e.
$\chi_m(M_{irr}^{AF})$ does not change. This leads to the
following expression of $A_L$:

\begin{subequations}
\begin{eqnarray}
&A_L(H,\psi)=\frac {I(2\omega,+H)-I(2\omega,-H)}
{I(2\omega,+H)+I(2\omega,-H)}\label{AL1}\\&= \frac {2
R_{L}^{rev}R_{L}^{irr} cos(\varphi_{rev}-\varphi_{irr})\ tan^2\psi
+ 2 R_{L}^{rev} tan\psi \cdot cos\varphi_{rev}} {1 +
({R_{L}^{rev}}^2 +{R_{L}^{irr}}^2) \ tan^2\psi + 2 R_{L}^{irr}
tan\psi \cdot cos\varphi_{irr}}\label{AL}
\end{eqnarray}
\end{subequations}
where $\psi$ is the analyzer angle and
$R_{L}^{rev}=|\frac{\chi_{m}(M_{rev}^{F/AF})}{\chi_{nm}}|_L$
(correspondingly, $R_{L}^{irr}
=|\frac{\chi_{m}(M_{irr}^{AF})}{\chi_{nm}}|_L$). Experimentally,
for each analyser angle $\psi$, the SH intensity is measured upon
reversal of the F layer. This enables us to determine $A_{L}
(\psi)$. Examples are shown in Fig.~\ref{fig3}.

Due to the irreversible term and according to Eq.~(\ref{AL}), two
distinct absolute extrema values are evidenced. The difference
between them depends on the bias field (it is lower for smaller
biases as shown in the inset). By fitting the asymmetry $A_L$ with
Eq.~(\ref{AL}), $R_{L}^{rev}$ and $R_{L}^{irr}$ are obtained.
Since the effective non-magnetic contribution to the
second-harmonic generation signal is constant~\cite{chinm},
$|\chi_m(M_{irr}^{AF})/\chi_{nm}|_L$ (respectively
$|\chi_{m}(M_{rev}^{F/AF})/\chi_{nm}|_L $) is directly
proportional to $M_{irr}^{AF}$ (respectively $M_{rev}^{F/AF}$)
along the magnetic field direction~\cite{sh}.

Our key experimental results (Fig.~\ref{fig2}c) are as follows: i)
the ion fluence dependence of the irreversible uncompensated spins
(full triangles) reproduces that of the exchange bias field
(Fig.~\ref{fig2}a), ii) a 13-fold increase in the irreversible
magnetization  and the one order of magnitude smaller interfacial
magnetic component enhancement (empty triangles) result in a bias
field enhancement by a factor of 1.8,  iii) the proportion of
irreversible (compared to reversible) spins given by the ratio
$\chi_m(M_{irr}^{AF})/\chi_{m}(M_{rev}^{F/AF})$ can reach about
10\%. In the same fluence regime, the linear Kerr rotation due to
the bulk F layer is constant~\cite{Mougin}. The fluence behavior
of $M_{irr}^{AF}$ and  $H_{eb}$ indicates that pinned
uncompensated spins in the AF layer control the bias field.
Moreover, this rigid AF moment is much smaller than the reversible
magnetization (maximum ca. 10 \%). According to the domain state
model~\cite{Nowak,Nowak2}, the exchange bias field is
 $H_{eb}= J_{int}M_{irr}^{AF}/\mu t$, with $J_{int}$ the
interface coupling, $\mu$ the magnetic moment per atom and $t$ the
F layer thickness. Random exchange defects or antisites are formed
in the AF layer via diluting impurities or irradiation. Under the
action of the exchange field originating from the magnetized F
layer, the AF domain structure is triggered and the uncompensated
spins associated to such defects exhibit an excess magnetization.
The latter is presumably reversible in the vicinity of the F layer
and pinned if deeper in the AF layer.  In the domain state model,
both  interface coupling and  magnetic moment are assumed to be
fixed and dilution only drives $M_{irr}^{AF}$. In our experimental
findings, additional information regarding the interfacial
contribution is obtained; changes in local chemical order due to
irradiation lead to a slight evolution of the interfacial
reversible magnetization. This mixing effect counteracts the
increase of $M_{irr}^{AF}$ and - for fluences larger than those
investigated here - reduces the bias as reported in
Ref~\cite{Mougin}.

To obtain insight on the exchange bias mechanism, we have combined
ion irradiation-induced tuning of the irreversible magnetization
in the AF layer with interface-selective second-harmonic
magneto-optic Kerr detection. Our main results do not rely on the
technique used to tune the uncompensated spins: whereas in
irradiated AF (as in diluted AF), the surplus magnetization is
linked to artificially introduced defects, this pinned component
may be associated with such natural structural imperfections as
grain boundaries in standard F/AF bilayers. Finally, we conclude
that the irreversible uncompensated AF spins (wherever they are
pinned) drive the exchange bias field.\\

\begin{acknowledgments}
The stay of L.C.S. at Orsay was financially supported by
CNPq/Brazil. B.H. acknowledges support by CNRS for a sabbatical
stay in Orsay. Partial support by the Deutsche
Forschungsgemeinschaft is gratefully acknowledged. T.M.
acknowledges support by the Studienstiftung des deutschen Volkes.

\end{acknowledgments}

\bibliography{AMOUGINresubmitLU8643}

\begin{thebibliography}{21}
\bibitem[{SHi()}]{SHintensity}
\bibinfo{note}{Experimentally, see Fig.~\ref{fig1}b, where the magnetic signal
  is 1.5\% of the average SH signal in transverse}.
\bibitem[{chi()}]{chinm}
\bibinfo{note}{The tensor elements entering $|\chi_{nm}|_T$ obtained from
  Fig.~\ref{fig1}b differ from those entering $|\chi_{nm}|_L$ but their dose
  dependence must be the same.}
\end{thebibliography}

\end{document}